\documentclass[a4paper]{jpconf}
\usepackage{graphicx}
\usepackage[numbers]{natbib}
\begin{document}
\title{Simulations of Magnetised Stellar-Wind Bubbles}

\author{J Mackey$^{1,2}$, S Green$^{1,2}$, M Moutzouri$^{1,2}$}

\address{$^1$ Dublin Institute for Advanced Studies, Astronomy \& Astrophysics Section, 31 Fitzwilliam Place, Dublin 2, Ireland}
\address{$^2$ Centre for Astroparticle Physics and Astrophysics (CAPPA), DIAS Dunsink Observatory, Dunsink Lane, Dublin 15, Ireland}

\ead{jmackey@cp.dias.ie}

\begin{abstract}
Initial results are presented from 3D MHD modelling of stellar-wind bubbles around O stars moving supersonically through the ISM.
We describe algorithm updates that enable high-resolution 3D MHD simulations at reasonable computational cost.
We apply the methods to the simulation of the astrosphere of a rotating massive star moving with 30\,km\,s$^{-1}$ through the diffuse interstellar medium, for two different stellar magnetic field strengths, 10\,G and 100\,G.
Features in the flow are described and compared with similar models for the Heliosphere.
The shocked interstellar medium becomes asymmetric with the inclusion of a magnetic field misaligned with the star's direction of motion, with observable consequences.
When the Alfv\'enic Mach number of the wind is $\leq10$ then the stellar magnetic field begins to affect the structure of the wind bubble and features related to the magnetic axis of the star become visible at parsec scales.
Prospects for predicting and measuring non-thermal radiation are discussed.
\end{abstract}

%%%%%%%%%%%%%%%%%%%%%%%%%%%%%%%%%
\section{Introduction}
%%%%%%%%%%%%%%%%%%%%%%%%%%%%%%%%%
There are many common physical processes and other similarities in the study of the Heliosphere and of the astrospheres of massive stars, but also a few differences.
Observations have shown the structure of the Solar Wind \citep{BalBadBon19, KasBalBel19} and Heliosphere \citep{McCAllBoc09, DiaKriMit17, ZanNakWeb19} in incredible detail, and global 3D computer models \citep{PogZanOgi06} have shown how these data can be interpreted in the context of the magnetohydrodynamics of partially ionized plasmas.
Much more limited observations are possible for other Sun-like stars, but we can measure their mass-loss rates and some properties of their stellar-wind bubbles \citep{WooMueZan05}.
Massive stars are much rarer than Sun-like stars and the nearest are $\sim10^2$\,pc distant \citep{GvaLanMac12}, but their extreme luminosity helps with observing their astrospheres.
Winds and extreme ultraviolet (EUV; $h\nu>13.6$\,eV) radiation from massive stars are many orders of magnitude stronger than for Solar-type stars \citep[e.g.][]{Lan12}, and their wind bubbles \citep{CasMcCWea75} and photoionized H~\textsc{ii} regions \citep{MacGvaMoh15} are of order parsec scale, $\sim10^3\times$ larger than the Heliosphere.
The mass of displaced interstellar material in the bow shock is correspondingly much larger, and this means that it is often easier to observe the astrosphere around a massive star than that around a Sun-like star.

Understanding the astrospheres of massive stars is more complicated than the Heliosphere because of the different timescales involved.
The size of the Heliosphere is $D\approx 100$\,A.U., and the Sun is moving with velocity $v_\star\sim10$\,km\,s$^{-1}$, and so a dynamical timescale $\tau_\mathrm{d}\equiv D/v_\star\sim10^2$\,years can be estimated.
This is many orders of magnitude shorter than the evolutionary timescale of the Sun ($\sim10^{10}$\,years), and shorter than the variation timescale for the local interstellar medium (ISM) properties encountered by the Heliosphere (i.e.~density fluctuations are weak on 100\,A.U. scales).
For massive stars, typical wind-bubble sizes are $\sim 1$\,pc (1\,pc$=3.086\times10^{18}$\,cm) and space velocities are again typically $v_\star\sim10$\,km\,s$^{-1}$, leading to $\tau_\mathrm{d}\sim10^5$\,years.
This is not too different from the stellar nuclear timescale over which a massive star evolves significantly ($\sim10^6$\,years, or shorter for helium-burning and later phases).
Furthermore the insterstellar medium (ISM) often has significant density fluctuations over the parsec length-scales traversed by a runaway massive star over $10^5$ years, and this means that the external ram pressure may change faster than the astrosphere can relax to its equilibrium size and shape.
Recent 2D simulations of the Bubble Nebula \citep{GreMacHaw19} modelled a star moving through a uniform ISM, and some synthetic observations such as predicted H$\alpha$ intensity maps of the nebula show discrepency with observations because the massive star BD+60$^{\circ}$\,2522 is apparently embedded in a medium of increasing density along its direction of motion.

Pioneering 2D hydrodynamical simulations \citep{GarLanMac96, ComKap98, BloKoe98} have shown the complexity of astrospheres from evolving and runaway stars in different environments.
These wind-blown bubbles have been studied for runaway massive stars of different masses  for their entire evolution \citep{MeyMacLan14}, and also the expansion of a supernova explosion into the pre-shaped circumstellar environment \citep{MeyLanMac15}.
Magnetic fields have been included in astrosphere simulations for massive hot stars \citep{vanMelMar15, MeyMigKui17}, but until recently only for 2D calculations with field geometry limited by the rotational symmetry.
A 3D implementation in spherical coordinates has recently been presented and applied to astrospheres of massive stars \citep{SchBaaFic20}.
The need for 3D simulations is twofold:
\begin{enumerate}
\item it allows us to use a general ISM magnetic field orientation that is not parallel to the direction of motion and/or rotation axis of the star being modelled; and
\item the forward shock is often radiative and highly compressed, subject to thin-shell instabilities \citep{DgaBurNor96}, which can lead to an artificial accumulation of dense gas along the symmetry axis in 2D calculations, compromising the validity of the solution \citep[e.g.][]{GreMacHaw19}.
\end{enumerate}

In this contribution we describe an upgrade of the \textsc{pion} magnetohydrodynamics (MHD) software \citep{MacLim11, Mac12} and some initial results of 3D MHD simulations of astrospheres.

\begin{figure}
\includegraphics[width=0.6\textwidth]{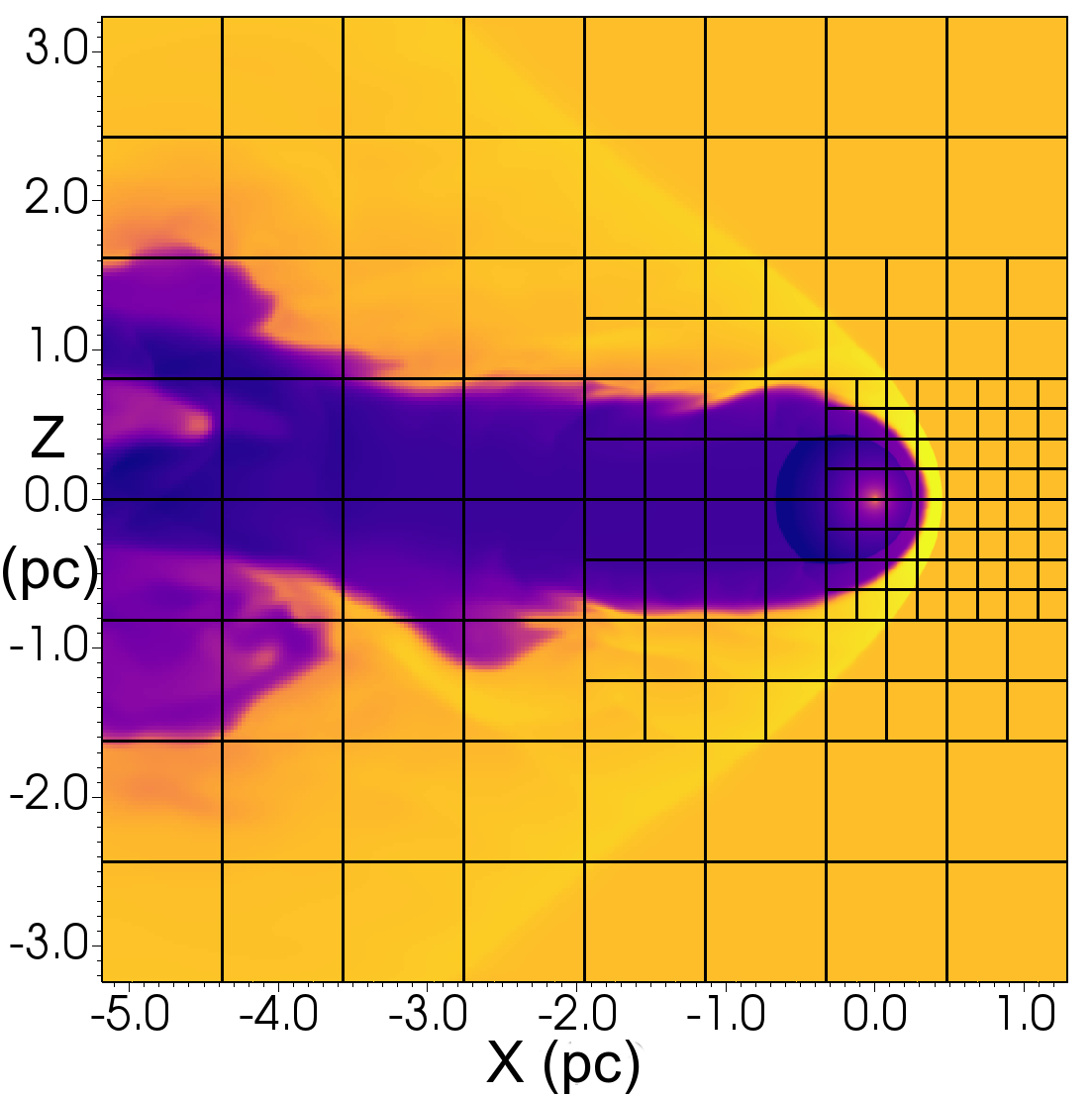}\hspace{2pc}%
\begin{minipage}[b]{0.38\textwidth}\caption{\label{fig:grid}} An example of the nested-grid configuration for modelling a stellar-wind bow shock.
The overlaid grid shows blocks of $32^3$ grid cells.
Log of gas density plotted for a 3D hydrodynamic simulation, run with 3 levels of refinement and $256^3$ grid cells at each level.
\end{minipage}
\end{figure}

%%%%%%%%%%%%%%%%%%%%%%%%%%%%%%%%%
\section{Methods}
%%%%%%%%%%%%%%%%%%%%%%%%%%%%%%%%%
\textsc{pion} is a finite-volume fluid-dynamics programme for solving the Euler \citep{MacLim10} and ideal-MHD \citep{MacLim11} equations on a rectilinear mesh, including a raytracer coupled to a chemical-kinetics solver to track the microphysical heating/cooling and the ionization of hydrogen by photoionization and collisional processes \citep{Mac12}.
We have recently added static mesh-refinement capabilities, to make 3D simulations of wind bubbles tractable with modest computing resources.
In brief, the improvements consist of the following:
\begin{enumerate}
\item
We implemented coarse-to-fine interpolation (prolongation) to populate the boundaries of refined grids with data from their parent grid, and fine-to-coarse averaging (restriction) to update regions of coarse grids with more accurately calculated data from an underlying finer-level grid \citep{TotRoe02}.
\item
We also implemented the boundary flux correction that ensures conservation of conserved quantities across different grid levels \citep{BerCol89}.
\item
We updated the raytracing algorithms for ionizing radiation from point sources so that they work on a multiply nested grid.
\item
We improved the Riemann solvers for HD and MHD, adding the robust (but diffusive) HLL solver for cases where the more accurate Roe and HLLD solvers do not maintain positive pressure or density \citep{MigZanTze12}.
\end{enumerate}

The method is similar to some previous 2D algorithms \citep{YorKai95, FreHenYor06}, using nested grids that differ in spatial and temporal resolution by a factor of 2 at each level.
For each dimension, the focus of the nested grids can be at the centre or the negative or positive extremity of the coarsest grid.
More details will be presented in a forthcoming paper, to accompany a public release of the software.
An example of a slice through the midplane of a 3D hydrodynamic simulation of a stellar-wind bow shock is shown in Fig.~\ref{fig:grid}, where the grid shows blocks of $32^3$ grid cells on three levels of refinement, focused on the apex of the bow shock.

\section{Results}
A 3D MHD simulation of a stellar-wind bow shock was set up with parameters given in Table~\ref{tab:3dmhd}.
The ISM values are typical of the diffuse gas in the Galactic plane within a factor of 2, \citep{MeyMacLan14} and the pressure is appropriate for photoionized gas at a temperature of $T\approx10^4$\,K.
The stellar values are typical for a massive star, here moving with velocity $v_\star=30\,\mathrm{km\,s}^{-1}$ through the ISM.
For these values, the standoff distance of the bow shock is
\begin{equation}
R_\mathrm{SO} \equiv \sqrt{\frac{\dot{M}v_\infty}{4\pi\rho_0 (v_\star*2+a^2)}} \approx0.60\,\mathrm{pc} \;,
\end{equation}
where symbols are as defined in Table~\ref{tab:3dmhd}, and $a\approx13.3\,$km\,s$^{-1}$ is the adiabatic sound speed in the photoionized ISM.
This is where we expect to find the wind termination-shock in the upstream direction.

\begin{table}
\caption{\label{tab:3dmhd}Parameters of the interstellar medium (ISM) and stellar wind for 3D MHD simulation of a bow shock.}
\begin{center}
\begin{tabular}{ll}
\br
Parameter&Value\\
\mr
ISM density, $\rho_0$ & $2.0\times10^{-24}\,\mathrm{g\,cm}^{-3}$ \\
ISM pressure, $p_g$   & $2.9\times10^{-12}\,\mathrm{dyne\,cm}^{-2}$ \\
ISM velocity, $\mathbf{v}$ & $[-30,0,0] \,\mathrm{km\,s}^{-1}$ \\
ISM B-field, $\mathbf{B}_0$  & $[4,1,1]\times10^{-6}$\,G \\
\mr
Wind mass-loss rate, $\dot{M}$ & $10^{-7}\,\mathrm{M}_\odot\,\mathrm{yr}^{-1}$ \\
Wind terminal velocity, $v_\infty$ & 1500\,km\,s$^{-1}$ \\
Surface rotation (equator), $v_\mathrm{rot}$  & 100\,km\,s$^{-1}$ \\
Surface split-monopole field strength, $|\mathbf{B}|$ & 10\,G \\
Surface temperature, $T_\mathrm{eff}$ & 35\,000\,K \\
\br
\end{tabular}
\end{center}
\end{table}

A simulation was initialised with a coarse grid of $128^3$ grid cells and volume $\approx8^3$ pc (each cell has diameter $\Delta x=0.0622$\,pc).
The simulation extents in the $x$-direction are $x\in[-6.30,1.67]$\,pc, and $\{y,z\}\in[-3.98,3.98]$\,pc.
The nested grids are centred on $[1.67,0,0]$\,pc, and the star is placed at the origin.
Two levels of refinement are added to the coarse grid, giving a finest level cell-diameter $\Delta x=0.0156$\,pc.
The wind inner boundary is initiated at at radius of 0.311\,pc, corresponding to 20 grid cells on the most refined level.
The same gas radiative heating and cooling prescription as in ref.~\citep{GreMacHaw19} was used, appropriate for photoionized gas with chemical abundances close to that of the Sun. 
A second-order-accurate integration scheme (in time and space) was used with the HLL Riemann solver to evolve the simulation.

This relatively low-resolution calculation takes approximately $3\times10^3$ core-hours to run to completion.
A higher resolution simulation with $256^3$ grid cells per level takes $\sim5\times10^4$ core-hours, and $512^3$ would take $\approx10^6$ core-hours.
The weak scaling of \textsc{pion} for these simulations is very good, and so it is feasible to run high-resolution calculations on many cores in a few days to a few weeks.

\begin{figure}
\includegraphics[width=1.0\linewidth]{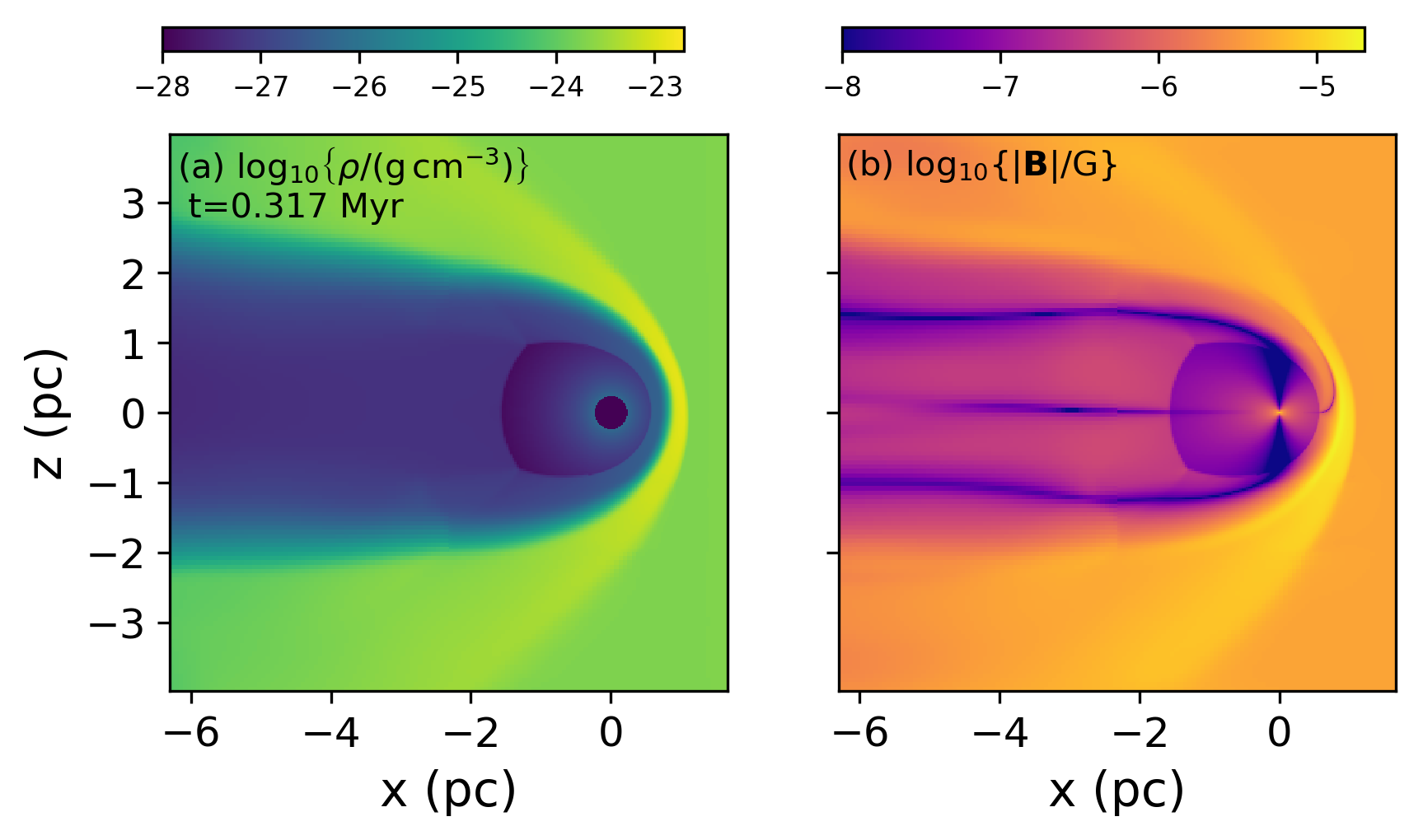}
\caption{\label{fig:b3d_DB} (a) Gas density, $\log_{10}\left\{\rho/\mathrm{(g\,cm}^{-3})\right\}$, and (b) magnetic field magnitude, $\log_{10}(|\mathbf{B}|/\mathrm{G})$, in the $x$-$z$ plane through $y=0$ are plotted on a logarithmic scale as indicated, for a 3D MHD simulation of a bow shock produced by a massive star.  The star is at the origin and moving in the $+\hat{x}$ direction.
The magnetic axis of the star is $\hat{z}$, the stellar surface field is $B=10$\,G, and the upstream ISM field is $\mathbf{B}_0=[4,1,1]\times10^{-6}$\,G.}
\end{figure}

Results at $t\approx0.32$\,Myr ($\tau_\mathrm{d}=16$ if we use $R_\mathrm{SO}$ as the size scale) are plotted in Fig.~\ref{fig:b3d_DB} for (a) $\rho$ and (b) $|\mathbf{B}|$ in the $x$-$z$ plane.
The wind termination-shock and contact discontinuity are easily discernable in both panels, as is the bow shock produced by the supersonic motion of the star through the ISM.
Already in Fig.~\ref{fig:b3d_DB} we can see morphological differences with respect to a hydrodynamic solution: the compression of the bow shock is larger in the upper half-plane than the lower, because of the orientation of the ISM magnetic field.
Without a magnetic field the solution should be axisymmetric, modulo instabilities that can form \citep{ComKap98}.
This would result in a brighter bow shock when observed in optical spectral lines such as H$\alpha$ and [O\,\textsc{iii}] (5007\AA), for which the emissivity is $\propto \rho^2$ in photoionized gas.
The wake behind the bow shock is also distorted by the ISM magnetic field and is no longer axisymmetric about the $x$-axis.
Both of these effects could introduce a systematic uncertainty in interpreting observations if one uses the symmetry axis of a bow shock to infer the direction of motion of the star.
The distortions get stronger as the interstellar magnetic field strength increases.

\begin{figure}
\includegraphics[width=1.0\linewidth]{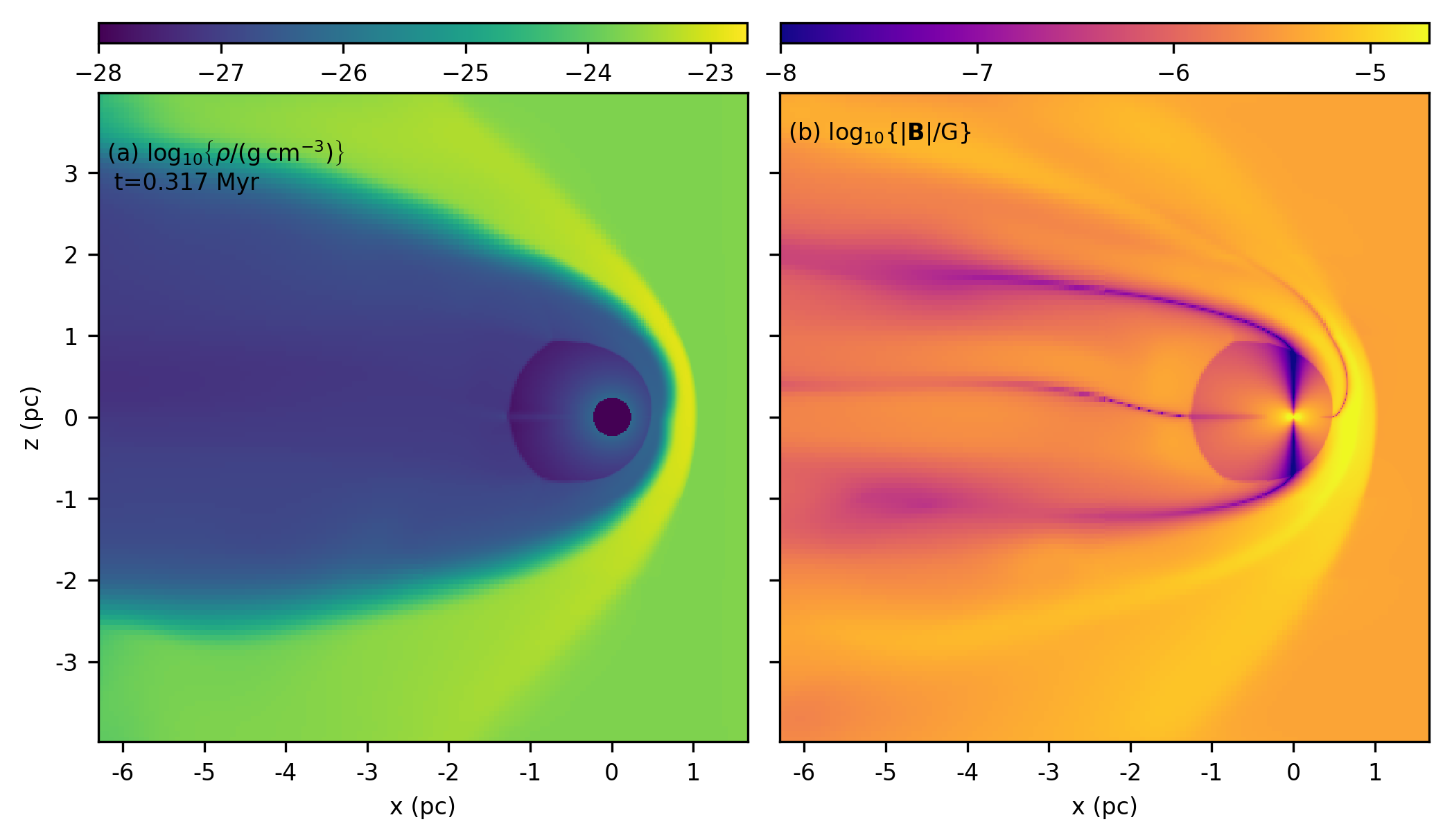}
\caption{\label{fig:b3d_B100} As Fig.~\ref{fig:b3d_DB} but for a simulation with a $10\times$ stronger stellar surface field of $B=100$\,G.}
\end{figure}

Fig.~\ref{fig:b3d_DB} (b) shows the magnetic structure of the astrosphere.
The Parker spiral from the rotating star ensures that $B\propto r^{-1}$ near the equatorial plane and $B\propto r^{-2}$ near the poles, with a current sheet at the equator across which the field lines switch from directed inwards to directed outwards.
The current sheet is preserved across the termination shock and is swept back (in the upper half-plane) along the contact discontinuity.
The very weakly magnetised regions emanating from the poles are also swept downstream in the shocked wind into the wake.
These features are consistent with MHD modelling of the Heliosphere \citep{PogZanOgi06} except that the length scales are $10^3\times$ larger.

The current sheet in the equatorial plane does not affect the structure of the wind bubble as long as the wind is only weakly magnetised, but if (keeping other parameters constant) the stellar surface field is increased to 100\,G (allowed by observational upper limits for most O stars \citep{FosCasSch15}) then some effects can be noticed in the shocked wind, shown in Fig.~\ref{fig:b3d_B100}.
Here, panel (a) shows an increased density in the equatorial plane in the freely-expanding wind (not present for stellar field of 10\,G), driven by the pressure gradient that arised from reconnection the current sheet and the associated very low magnetic pressure.
The contact discontinuity also moves closer to the star just below the equator in the upstream direction.
Both of these effects become more pronounced if the stellar surface magnetic field is increased further.
This is possibly related to the numerical issue in Heliosphere simulations, where a V-shaped structure forms on the contact discontinuity in ideal-MHD simulations \citep{WasTan01}.
This feature was shown to be strongly dependent on the numerical resolution at the current sheet, and disappeared with the inclusion of neutrals as a separate fluid \citep{PogZanOgi06,WasZanHu15}.

Fig.~\ref{fig:b3d_B100} (b) also shows that the magnetic field in the shocked wind bubble is about the same strength as the ISM field, and there is no sharp change in the field strength across the contact discontinuity in the upstream direction.
Assuming that the wind termination-shocks of massive stars are reasonably efficient at accelerating electrons, this could have observable consequences.
2D hydrodynamic simulations with an analytically estimated magnetic field strength have been used to predict synchrotron radiation from bow shocks \citep{DelPoh18}.
Their calculations show that the relativistic electrons are well-confined in the wind bubble and so the synchrotron radiation is dominated by emission from within the wind bubble if the wind has comparable magnetic field to the shocked ISM.
This is in strong contrast with Bremsstrahlung which is dominated (at radio frequencies) by the photoionized and shocked ISM, external to the wind bubble, because the thermal electron density is orders of magnitude larger in the shocked ISM than in the wind bubble.
Sensitive radio observations could disentangle these two components and constrain the stellar surface magnetic field and the particle acceleration efficiency \citep{BenRomMar10, DelBosMul18}.
Application of the method of \citep{DelPoh18} to 3D MHD simulations, where the magnetic field is calculated self-consistently with the fluid dynamics, is an interesting avenue for making more detailed predictions of non-thermal radiation.

\section{Outlook}
We have presented some initial results from 3D MHD modelling of stellar-wind bubbles around O stars moving supersonically through the ISM.
Algorithm updates have also been briefly described which enable high-resolution 3D MHD simulations of the astrospheres of massive stars at reasonable computational cost.
A paper describing the algorithms and tests is currently in preparation, and we are beginning to apply the methods to the astrospheres of some well-known runaway stars, e.g., BD+60$^{\circ}$\,2522 and the Bubble Nebula that surrounds it \citep{GreMacHaw19}, and BD+43$^{\circ}$\,3654 and its parsec-scale bow shock \citep{BenRomMar10}.

For a stellar magnetic field strength of 10\,G and wind properties given in Table~\ref{tab:3dmhd}, the stellar magnetic field does not significantly affect the structure of the astrosphere, whereas the interstellar field does change the morphology of the bow shock to some extent.
For this case the Alfv\'enic Mach number of the wind is $\mathcal{M}_\mathrm{A}\approx70$, and so the magnetic pressure is significantly less than the ram pressure in the unshocked wind and less than the thermal pressure in the shocked wind.
The interstellar $\mathcal{M}_\mathrm{A}\approx3.5$ and so the magnetic effects on the shocked ISM are correspondingly more evident.

With a 100\,G stellar field, the Alfv\'enic Mach number of the wind is only $\mathcal{M}_\mathrm{A}\approx7$, and so the magnetic pressure in the shocked wind is signficant but still smaller than the thermal pressure.
The Axford-Cranfill effect means that the importance of magnetic pressure increases outwards in the shocked wind; for a review see \cite{Zan99} and for a recent application to wind bubbles of massive stars see \cite{ZirPtu18}.
This introduces small changes to the shape of the wind bubble, especially near the magnetic equator at the contact discontinuity.
An even stronger stellar field would be expected to produce a wind bubble that is signficantly affected by magnetic effects.
In that case the boundary condition that we impose is probably not appropriate anyway and a more complicated wind injection method would be required.
Nevertheless, our results show that for $\mathcal{M}_\mathrm{A}\geq10$ the effects of the equatorial current sheet on the flow dynamics are modest, and decreasing as $\mathcal{M}_\mathrm{A}$ increases.
The ideal MHD algorithms that are presented here are therefore adequate to describe astrospheres from the majority of massive stars.

3D MHD simulations have significantly more predictive power than hydrodynamic calculations because (with some assumptions) the non-thermal radiation from the nebula can be calculated more realistically.
This holds the promise of constraining the efficiency of particle acceleration in stellar-wind shocks, a topic of considerable interest \citep{BenRomMar10, DelBosMul18, HESS2018_Bowshocks}  for current and upcoming observing facilities.

\ack
JM acknowledges funding from a Royal Society-SFI University Research Fellowship (14/RS-URF/3219).
SG is funded by a Hamilton Scholarship from the Dublin Institute for Advanced Studies.
MM acknowledges funding from a Royal Society Research Fellows Enhancement Award (RGF\textbackslash EA\textbackslash 180214).
We acknowledge the SFI/HEA Irish Centre for High-End Computing (ICHEC) for the provision of computational facilities and support (project dsast022c).
%Figures were generated using (a) python with matplotlib \citep{Hun07} and SciPy/Numpy \citep{VirGomOli20} and (b) VisIt \citep{HPV:VisIt}; VisIt is supported by the Department of Energy with funding from the Advanced Simulation and Computing Program and the Scientific Discovery through Advanced Computing Program

\bibliographystyle{iopart-num}
\bibliography{./refs}

\end{document}